\documentclass[final]{aipproc}
\layoutstyle{6x9}
\newcommand{\be}{\begin{equation}}    
\newcommand{\ee}{\end{equation}}
\newcommand{\beq}{\begin{eqnarray}}
\newcommand{\eeq}{\end{eqnarray}}
\newcommand{\beqn}{\begin{eqnarray*}}
\newcommand{\eeqn}{\end{eqnarray*}}

\def\aprb{{\rm APR2~}}
\def\apb{{\rm APRB200~}}
\def\apc{{\rm APRB120~}}
\def\fa{{\rm BBS1~}}
\def\fb{{\rm BBS2~}}
\def\gr{{\rm G240~}}
\def\sqm2{{\rm SS0~}}
\def\sqm1{{\rm SS1~}}
\def\scrust{{\rm SS2~}}
\def\lsim{\mathrel{\rlap{\lower2.5pt\hbox{\hskip1pt$\sim$}}
    \raise1pt\hbox{$<$}}}         
\def\gsim{\mathrel{\rlap{\lower2.5pt\hbox{\hskip1pt$\sim$}}
    \raise1pt\hbox{$>$}}}         

\begin{document}

\title{Gravitational waves from neutron stars described by modern EOS}

\author{O. Benhar}{
  address={ Dipartimento di Fisica ``G. Marconi",
Universit\'a degli Studi di Roma, ``La Sapienza",\\ P.le A. Moro
2, 00185 Roma, Italy},
altaddress={INFN, Sezione Roma 1,P.le A. Moro
2, 00185 Roma, Italy}
}

\author{V. Ferrari}{
  address={ Dipartimento di Fisica ``G. Marconi",
Universit\'a degli Studi di Roma, ``La Sapienza",\\ P.le A. Moro
2, 00185 Roma, Italy},
altaddress={INFN, Sezione Roma 1,P.le A. Moro
2, 00185 Roma, Italy}
}

\author{L. Gualtieri}{
  address={ Dipartimento di Fisica ``G. Marconi",
Universit\'a degli Studi di Roma, ``La Sapienza",\\ P.le A. Moro
2, 00185 Roma, Italy},
altaddress={INFN, Sezione Roma 1,P.le A. Moro
2, 00185 Roma, Italy}
}

\begin{abstract}
The frequencies and damping times of neutron star (and quark star)
oscillations have been computed using the most recent equations of
state available in the literature. We find that some of the empirical
relations that connect the frequencies and damping times of the modes to
the mass and radius of the star, and that
were previously derived in the literature need to be modified.
\end{abstract}

\maketitle

Asteroseismology, that is, the study of stellar properties through the
analysis of the proper oscillation frequencies, is a very useful tool.
For instance, it has been succesfully applied to study the internal composition of the Sun.

The oscillation modes of compact stars, like neutron stars (NS) or quark stars,
give a clear signature in the spectrum of the gravitational waves that these stars may emit in
several astrophysical processes.
It should be stressed that, according to General Relativity  the
modes of compact stars are not normal modes, because they
are damped by gravitational wave  (GW) emission; 
for this reason they are called ``quasi-normal modes" (QNM), with
characteristic frequencies and damping times,  that
carry information on the structure of the star and on the
behaviour of nuclear matter in the interior.  The study of QNM 
is called {\em gravitational wave asteroseismology} \cite{AK}.

Some years ago, Andersson and Kokkotas computed the frequencies and
damping times of the most relevant oscillation modes \cite{AK} (that
is, the modes that most likely would be excited by a perturbing event)
of a non rotating NS for a number of equations of state (EOS)
available at that time. They fitted the data with appropriate
functions of the radius and the mass of the star, showing how
these empirical relations could be used to put constraints on these
parameters if the frequency of one or more modes could be identified
in a detected gravitational signal. Knowing the mass and the radius,
we would gain information on the behaviour of matter in a NS core, 
at density that cannot be reproduced in a laboratory.

In recent years, a number of new EOS have been proposed to describe
matter at supranuclear densities, some of them allowing for the
formation of a core of strange baryons and/or deconfined quarks.  In
ref. \cite{paper}, that we summarize here, we have verified
whether, in the light of the recent developments, the empirical
relations derived in \cite{AK} are still appropriate or need to be
updated.

We have considered a variety of EOS.  For any of them we have obtained
the equilibrium configurations for assigned values of the mass, and 
solved the equations of stellar perturbations computing the
frequencies and damping times of the QNM. Then, we have fitted our
data with suitable functions of M and R to see whether the fits agree
with those of \cite{AK}.  We have extended the results of
\cite{AK} in two respects: we have considered more recent EOS, and we
have studied a larger set of QNM.


A NS is believed to be composed mainly by three different layers of different
composition: an outer crust, composed by heavy nuclei and free
electrons; an inner crust, composed by heavy muclei, free electrons
and neutrons; a core, composed by leptons, nucleons, and, in some
models, also hyperons or quarks.  There is an overall agreement on the
EOS describing the crust \cite{BPS,PRL}, while the composition of the
core is poorly known, due to the present limited understanding of
hadronic interactions. We have modeled the crust as in
\cite{BPS,PRL}, and used  various models for the
 matter in the core, which we summarize in Table \ref{table1}.

\begin{table}[h]
\begin{tabular}{ll}
\aprb & Akmal, Pandharipande, Ravenhall \cite{AP,APR} \\
\apb, \apc & \aprb \cite{AP,APR} + quark inner core \cite{MITB},
\cite{BR} \\
\fa & Baldo, Burgio, Schultze without hyperons \cite{BBS200}\\
\fb & Baldo, Burgio, Schultze with hyperons \cite{BBS200}\\
\gr & Glendenning, mean field approximation \cite{Gbook}\\
\end{tabular}
\caption{EOS included in our study}
\label{table1}
\end{table}
In addition  to the  above models we  have considered  the possibility
that  a star entirely  made of  quarks (strange  star) may  form.  The
models  denoted $\sqm1$  and  $\scrust$ correspond  to  a quark  star,
described by the MIT bag model \cite{MITB}, with or without a crust.

In order to find the frequencies and damping times of the quasi-normal
modes, we have solved the equations describing non radial
perturbations of a non rotating star in general relativity
\cite{lindet,CF1991}. In \cite{paper} we have considered several
oscillation modes. Here we only report the results on
the fundamental mode (f-mode), which gives the major contribution to GW emission.

\begin{figure}
  \includegraphics[height=.4\textheight,angle=270]{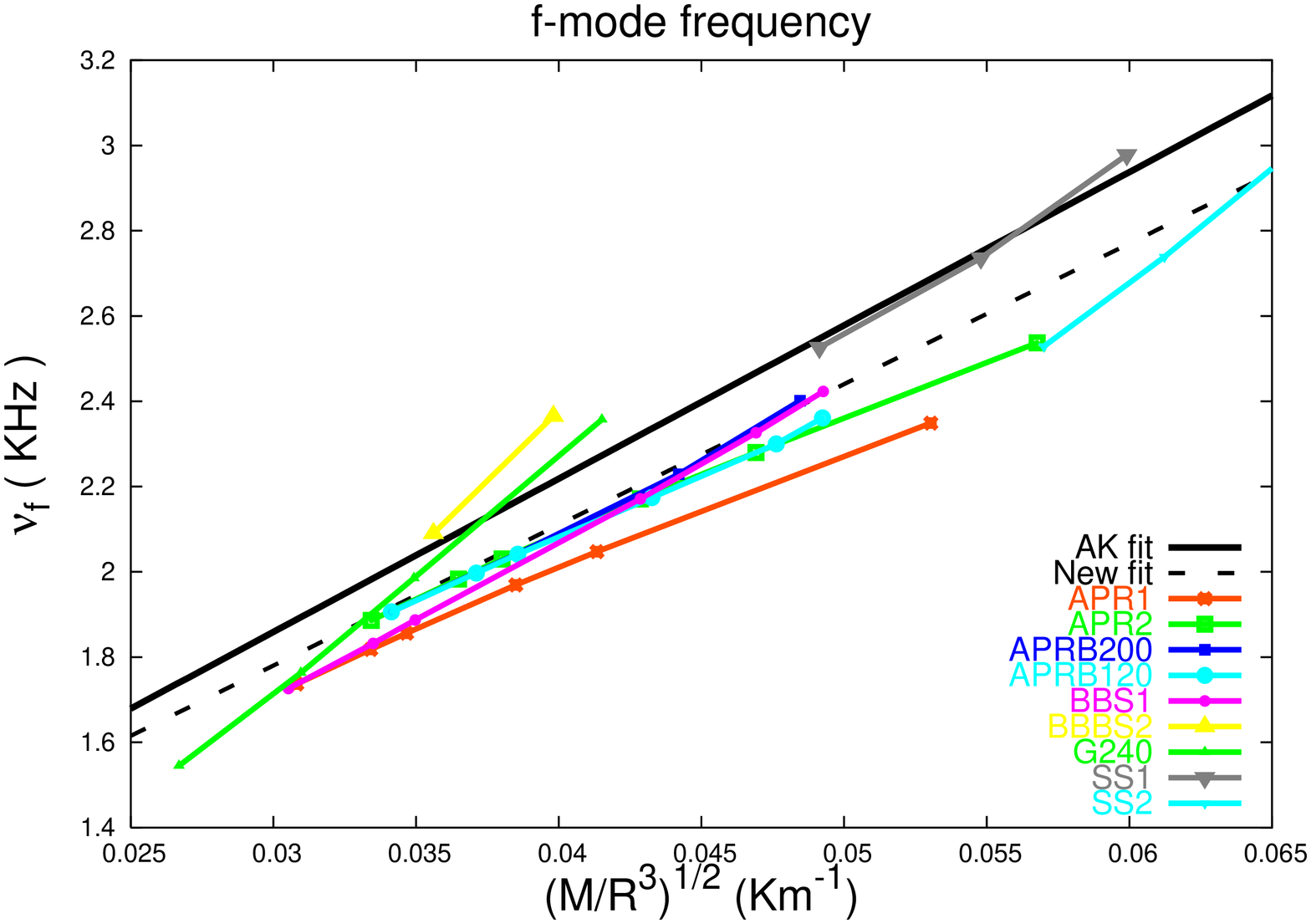}
  \includegraphics[height=.4\textheight,angle=270]{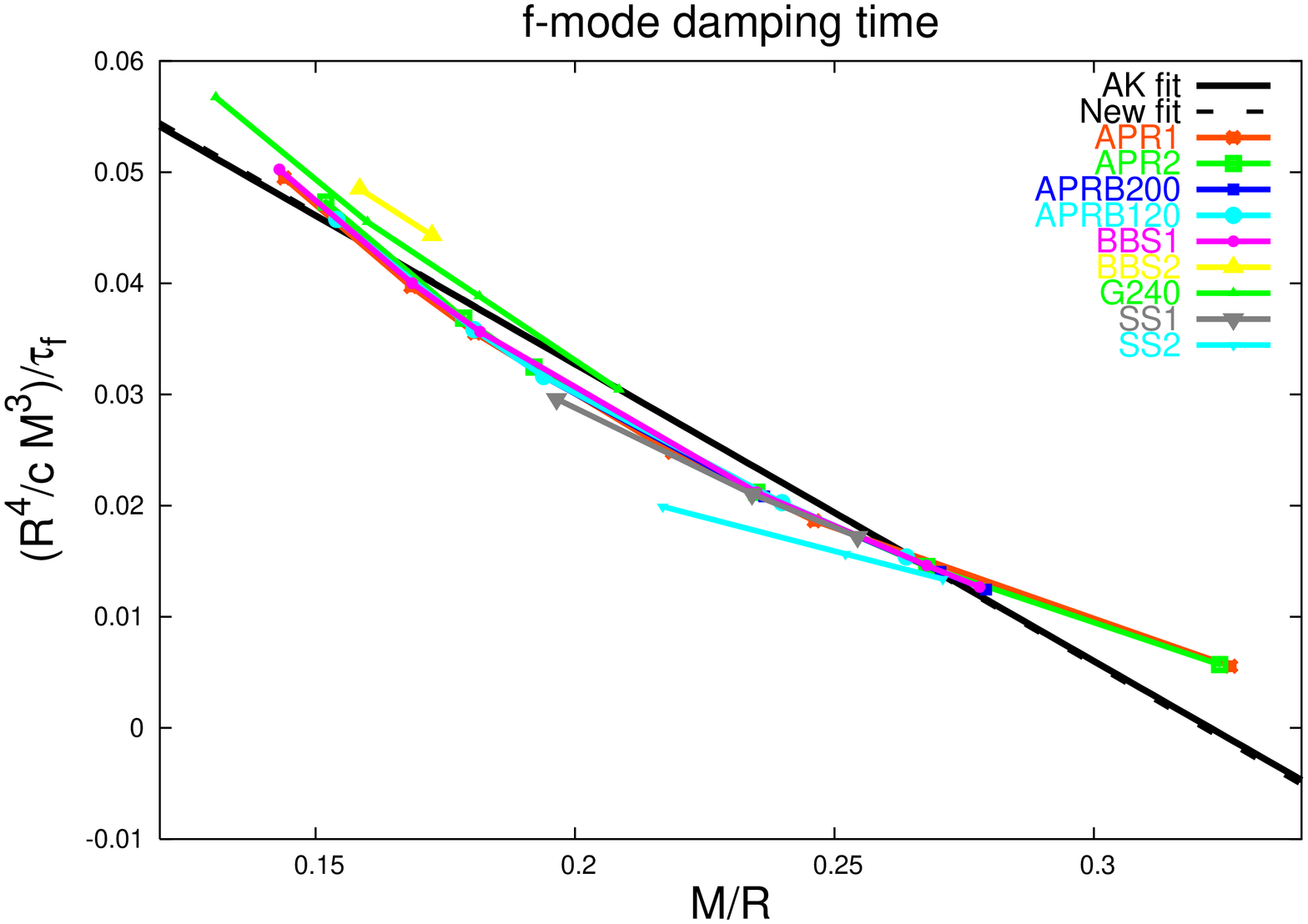}
\caption{The frequency of the fundamental mode is plotted in the left
panel as a function of the square root of the average density for the
different EOS considered in this paper.  The new fit is systematically
lower (about 100 Hz) than that previously derived in the literature. The damping time of the
fundamental mode is plotted in the right panel as a function of the
compactness $M/R$.}
\label{nuf}
\end{figure}

The data we derived can be fitted by the following expressions:
\be
\label{fitf}
\nu_f=a+b\sqrt{\frac{M}{R^3}}\,,~~~~~
\tau_f=\frac{R^4}{cM^3}\left[c+d\frac{M}{R}\right]^{-1}
\ee
with $a=0.79\pm0.09$ kHz, $b=33\pm2$ km kHz, $c=[8.7\pm0.2]\cdot
10^{-2}$ and $d=-0.271\pm0.009$.  Frequencies are expressed in kHz, 
masses and radii in km, damping times in s.  The data for the f-mode
and the fits are shown in Fig. \ref{nuf}.  In the left panel we plot
$\nu_f$ versus $\sqrt{M/R^3}$, for all considered stellar models.  Our
fit (\ref{fitf}) is plotted as a thick solid line, and the fit computed by Andersson and
Kokkotas in \cite{AK}, which is based on the EOSs considered in that paper, is
plotted as a dashed line labelled as `AK-fit'.  In the right panel we
plot the damping time $\tau_f$ versus the compactness $M/R$, our fit
and the corresponding AK-fit.  We can see that our new fit for $\nu_f$
is sistematically lower than the AK fit by about 100 Hz. Conversely,
our fit for the damping time is very similar to that found in
\cite{AK}.
The empirical relations derived above can be used, as described in \cite{AK,AK1},
to determine the mass and the radius of the star from the knowledge of the frequency and 
damping time of  the modes.

By comparing our results with the sensitivity curves of existing
gravitational detectors, we have shown in \cite{paper} that it is
unlikely that the first generation of interferometric antennas will
detect the GW emitted by an oscillating neutron star.  However, new
detectors are under investigation that should be much more sensitive
at frequencies above 1-2 kHz and that would be more appropriate to
detect these signals.  If the frequencies of the modes will be
identified in a detected signal, the simultaneous knowledge of the
mass of the emitting star will be crucial to understand its internal
composition.

\end{document}